\documentclass[letterpaper,reprint,amsmath,amssymb,aps,prl,twocolumn,showpacs,superscriptaddress]{revtex4-1}
\usepackage{graphicx}	
\begin{document}

\title{Distinguishing bulk and surface electron-phonon coupling in the topological insulator Bi$_2$Se$_3$ using time-resolved photoemission spectroscopy}
\author{J.~A. Sobota}
\affiliation{Stanford Institute for Materials and Energy Sciences, SLAC National Accelerator Laboratory, 2575 Sand Hill Road, Menlo Park, CA 94025, USA}
\affiliation{Geballe Laboratory for Advanced Materials, Department of Applied Physics, Stanford University, Stanford, CA 94305, USA}
\affiliation{Advanced Light Source, Lawrence Berkeley National Laboratory, Berkeley, California 94720, USA}
\author{S.-L. Yang}
\affiliation{Stanford Institute for Materials and Energy Sciences, SLAC National Accelerator Laboratory, 2575 Sand Hill Road, Menlo Park, CA 94025, USA}
\affiliation{Geballe Laboratory for Advanced Materials, Department of Applied Physics, Stanford University, Stanford, CA 94305, USA}
\affiliation{Department of Physics, Stanford University, Stanford, CA 94305, USA}
\author{D. Leuenberger}
\affiliation{Stanford Institute for Materials and Energy Sciences, SLAC National Accelerator Laboratory, 2575 Sand Hill Road, Menlo Park, CA 94025, USA}
\affiliation{Geballe Laboratory for Advanced Materials, Department of Applied Physics, Stanford University, Stanford, CA 94305, USA}
\author{A.~F. Kemper}
\affiliation{Computational Research Division, Lawrence Berkeley National Lab, 1 Cyclotron Road, Berkeley, CA 94720, USA}
\author{J.~G. Analytis}
\affiliation{Department of Physics, University of California, Berkeley, California 94720, USA}
\author{I.~R. Fisher}
\affiliation{Stanford Institute for Materials and Energy Sciences, SLAC National Accelerator Laboratory, 2575 Sand Hill Road, Menlo Park, CA 94025, USA}
\affiliation{Geballe Laboratory for Advanced Materials, Department of Applied Physics, Stanford University, Stanford, CA 94305, USA}
\author{P.~S. Kirchmann}
\email{kirchman@stanford.edu}
\affiliation{Stanford Institute for Materials and Energy Sciences, SLAC National Accelerator Laboratory, 2575 Sand Hill Road, Menlo Park, CA 94025, USA}
\author{T.~P. Devereaux}
\affiliation{Stanford Institute for Materials and Energy Sciences, SLAC National Accelerator Laboratory, 2575 Sand Hill Road, Menlo Park, CA 94025, USA}
\affiliation{Geballe Laboratory for Advanced Materials, Department of Applied Physics, Stanford University, Stanford, CA 94305, USA}
\author{Z.-X. Shen}
\email{zxshen@stanford.edu}
\affiliation{Stanford Institute for Materials and Energy Sciences, SLAC National Accelerator Laboratory, 2575 Sand Hill Road, Menlo Park, CA 94025, USA}
\affiliation{Geballe Laboratory for Advanced Materials, Department of Applied Physics, Stanford University, Stanford, CA 94305, USA}
\affiliation{Department of Physics, Stanford University, Stanford, CA 94305, USA}

\date{\today}

\begin{abstract}

We report time- and angle-resolved photoemission spectroscopy  measurements on the topological insulator Bi$_2$Se$_3$.  We observe oscillatory modulations of the electronic structure of both the bulk and surface states at a frequency of 2.23~THz due to coherent excitation of an A$_\textrm{1g}$ phonon mode. A distinct, additional frequency of 2.05~THz is observed in the surface state only.  The lower phonon frequency at the surface is attributed to the termination of the crystal and thus reduction of interlayer van der Waals forces, which serve as restorative forces for out-of-plane lattice distortions.  DFT calculations quantitatively reproduce the magnitude of the surface phonon softening.  These results represent the first band-resolved evidence of the A$_\textrm{1g}$ phonon mode coupling to the surface state in a topological insulator.
\end{abstract}
\pacs{78.47.J-, 73.20.-r, 63.20.kd, 79.60.-i}
\maketitle

Topological insulators (TIs) are materials that behave as electronic insulators in their bulks, but have robust  surface states (SSs) which enable metallic conduction \cite{Fu2007,Zhang2009,Chen2009,Xia2009,Qi2011}.  A particularly exciting property of the SS is that it is strongly spin-polarized, with the electrons' spin-orientations locked perpendicular to their momenta \cite{Hsieh2009a,Hsieh2009b}.  While this novel spin texture greatly reduces the phase space for spin-conserving scattering events \cite{Roushan2009,Zhang2009a}, there still remain scattering processes which give the SS electrons a finite lifetime and limit their ballistic transport, and thus must be considered for device applications \cite{Butch2010}.

Among these scattering processes, those driven by electron-phonon coupling (EPC) in particular have been the subject of intense study because they affect any finite-temperature application of the TIs.  The fundamental questions regarding EPC are: Which electronic states are involved, and to which phonon modes do they couple?  A number of recent measurements including helium atom scattering \cite{Zhu2011,Zhu2012,Howard2013} and inelastic transport \cite{Costache2014} seem to be arriving at a consensus that scattering in Bi$_2$Se$_3$ is dominated by a $\sim7 - 8$~meV optical A$_\textrm{1g}$ phonon  mode \cite{Richter1977}.  However, because of the coexistence of bulk and surface carriers in Bi$_2$Se$_3$ \cite{Analytis2010}, it is not clear from these experiments whether the measured  A$_\textrm{1g}$ mode coupling corresponds to EPC in the bulk or surface states.  In principle, this could be investigated by angle-resolved photoemission spectroscopy (ARPES) because of its capability to measure the electron self-energy directly on the SS band.   However, the ARPES results reported in the literature have been scattered: two works identified no particular mode coupling to the SS \cite{Hatch2011, Pan2012}, one identified an $\sim18$~meV mode \cite{Chen2013}, while yet another identified modes at both $\sim$3 and $\sim$18~meV \cite{Kondo2013}.  To-date no measurement has directly observed the A$_\textrm{1g}$ mode coupling to the SS band.

The discrepancy in ARPES measurements is likely attributed to the fact that the coupling occurs on a small energy scale accompanied by a weak spectral signature, so very high energy resolution is required to unambiguously detect it \cite{Howard2013}.  Here we circumvent this experimental difficulty by taking a complementary approach: rather than look for spectral signatures of EPC in the energy-domain using ARPES, we study EPC in the time-domain using ultrafast time-resolved ARPES (trARPES).  trARPES has already been extensively employed to study EPC in TIs, and in particular has elucidated the significant role of  the phonon-mediated interaction between the bulk and surface states \cite{Hajlaoui2012,Hajlaoui2013,Sobota2012,Wang2012a,Crepaldi2012,Crepaldi2013,Sobota2014}.   However, in these experiments EPC was studied via its role in electron population relaxation, which does not permit direct identification of the frequencies of the relevant phonon modes.  In contrast,  time-resolved reflectivity (TRR) experiments have observed coherent oscillations attributed to A$_\textrm{1g}$ optical phonons \cite{Qi2010, Kumar2011, Chen2012,Norimatsu2013a, Glinka2013}.  But because TRR lacks the energy- and momentum- resolution needed to distinguish electronic bands, it is not known whether these oscillations involve the TI SS.  The importance of checking this is exemplified by elemental bismuth:  it was recently shown by trARPES of Bi(111) that the  A$_\textrm{1g}$ oscillations measured in TRR arise predominantly from the bulk electrons rather than from the SS \cite{Faure2013}.

In this Letter we report a trARPES experiment on the TI Bi$_2$Se$_3$ which reveals coherent A$_\textrm{1g}$ optical phonons coupling to both bulk and surface bands.  The SS dispersion oscillates with two distinct frequencies of 2.05 and 2.22~THz, while the bulk conduction band (CB) oscillates with a single frequency of 2.23~THz.   The frequency common to both the SS and CB is associated with the bulk frequency of the A$_\textrm{1g}$ mode, while the additional frequency present in the SS is attributed to a softening of the phonon mode at the crystal surface.   We quantitatively reproduce this softening in density functional theory (DFT) calculations with frozen A$_\textrm{1g}$ phonon distortions.  These results represent the first direct evidence of  A$_\textrm{1g}$ phonon mode coupling to the SS band in a TI.

\begin{figure}
\resizebox{\columnwidth}{!}{\includegraphics{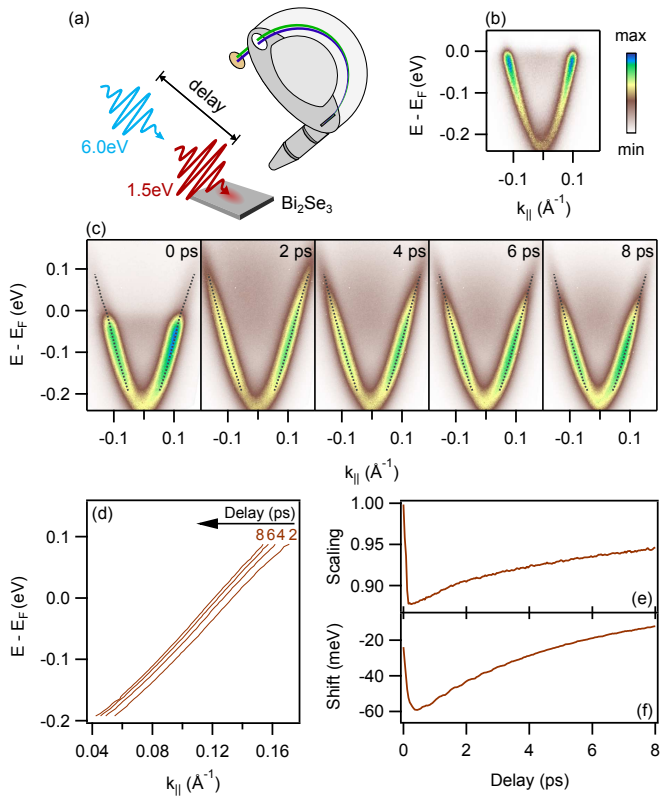}}
\caption{(a) Schematic of the experimental setup including 1.5~eV pump and 6~eV probe pulses. (b) ARPES spectrum of the sample at equilibrium (unpumped) at 40~K.  (c) trARPES spectra after optical excitation.  The dotted lines represent the SS dispersion extracted by MDC fitting.  (d) Zoom-in on the right-hand branch of the extracted dispersions, showing a time-dependent renormalization of the band.  (e) The time-dependent scaling and (f) shift of the SS dispersion.   
\label{RawCuts}}
\end{figure}

Our trARPES setup and methods of optimization are discussed in detail elsewhere \cite{Sobota2012,Yang2013}.  For this experiment we use a Coherent RegA Ti:Sapphire amplified laser operating at 100~kHz repetition rate.  We optically excite the sample using pulses of duration $\sim$50~fs and photon energy 1.5~eV,  and subsequently probe the electronic structure by photoemitting electrons with $\sim$150~fs, 6~eV pulses, as shown in Fig. \ref{RawCuts}(a).  For this experiment the incident pump fluence was 1.6~mJ/cm$^2$.  The photoelectrons are collected by a hemispherical electron analyzer.  The total energy resolution is $\sim$22~meV.    The photoemission measurements were performed on freshly cleaved samples at a temperature of 40~K in an ultrahigh vacuum chamber with pressure $<1\times10^{-10}$~torr.  Single crystals of Bi$_2$Se$_3$ were synthesized using conventional methods \cite{Analytis2010,Fisher2012}.    The band structure and frozen phonon calculations were performed using the Quantum Espresso \cite{Giannozzi2009} software package using the projector augmented wave method \cite{Kresse1993} and fully relativistic PBE exchange functionals \cite{Perdew1996}.  The wavefunction and density energy cutoffs were chosen to be 40~Ry and 450~Ry, respectively.

\begin{figure*}
\resizebox{1.8\columnwidth}{!}{\includegraphics{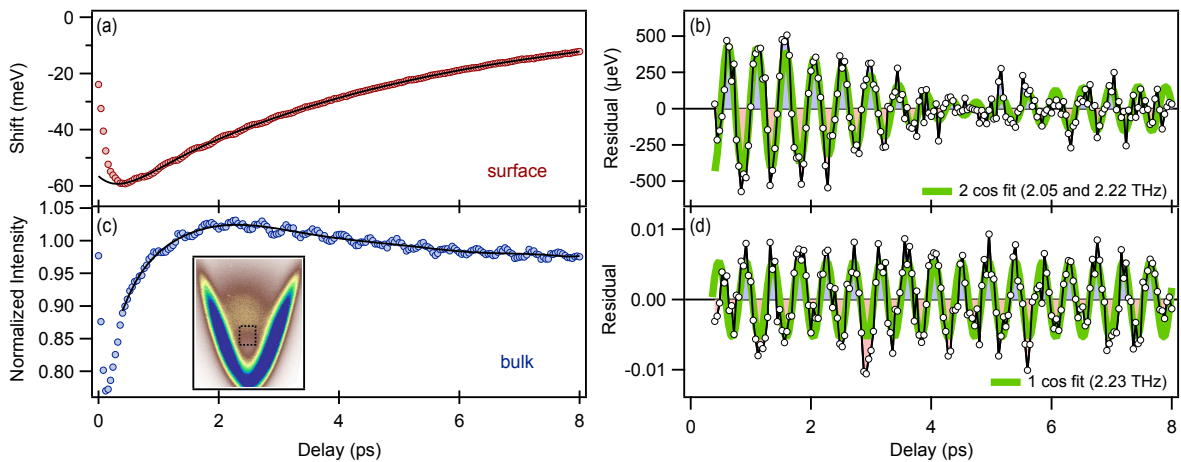}}
\caption{Oscillations in the SS (top) and CB (bottom).  (a) The time-dependent shift of the SS and a smooth curve fit (10$^\textrm{th}$ order polynomial). (b) Residuals from the fit.  The oscillations can be fit as a beating pattern due to the sum of two cosine functions (see text for fit details). (c) Photoemission intensity at the CB edge obtained by integrating within the window indicated in the inset.  The color scale in the inset is oversaturated to emphasize the CB.  The smooth curve fit is a 10$^\textrm{th}$ order polynomial.  (d) Residuals from the fit.  The oscillations can be fit with a single undamped cosine.
\label{oscillations}}
\end{figure*}

A representative ARPES spectrum  from Bi$_2$Se$_3$ at equilibrium at 40~K is shown in Fig. \ref{RawCuts}(b).  Fig. \ref{RawCuts}(c) presents trARPES data at select pump-probe delays.  At time zero the 1.5~eV pulse drives a direct transition from the valence band to high-lying unoccupied bulk and surface states \cite{Sobota2013}, and after $\sim$1~ps these electrons scatter down to the SS and CB \cite{Sobota2012}.  We begin our analysis by extracting the time-dependent dispersion of the SS band obtained via momentum-distribution curve (MDC) fitting \cite{SOM}. The dispersions are shown as dotted lines in Fig. \ref{RawCuts}(c) and plotted together in (d) (only the right-hand branch is shown for greater clarity). To parametrize the time-dependence of the dispersion, it is helpful to express the dispersion $\varepsilon(k,t)$ at delay $t$ in terms of an unperturbed dispersion $\varepsilon_0(k)$.  The simplest phenomenological model we found to fit the data was:

\begin{equation}
\varepsilon(k,t) = A(t) \cdot \varepsilon_0(k) + B(t)
\end{equation}

In other words, the dispersion at $t$ is scaled by $A(t)$ and shifted by $B(t)$ with respect to the unperturbed dispersion.  We used a self-consistent fitting procedure described in Ref. \cite{SOM} to  extract $\varepsilon_0(k)$, $A(t)$, and $B(t)$. The resulting scaling and shift parameters are shown in Figs. \ref{RawCuts}(e) and (f).  Note that we used $E_F$ as the energy reference for scaling; the scaling parameter is independent of this choice, but the values of the shift parameter depend on the energy reference used.  Both curves exhibit similar behavior: a rapid change around time zero followed by a return to equilibrium on a $>$5~ps timescale.

There is a fine structure to the time-dependent shift which is not apparent in Fig. \ref{RawCuts}(f).  In Fig. \ref{oscillations}(a) we re-plot the data, fit a smooth background (10$^\textrm{th}$ order polynomial) to the curve, and plot the corresponding residual in Fig. \ref{oscillations}(b). From this treatment it is clear that there is an oscillatory time dependence.  The oscillations have a magnitude $<500$~$\mu$eV which reaches a minimum at $\sim$4.5~ps and subsequently recovers.  This behavior can be understood as a beating pattern between two similar yet distinct oscillation frequencies.  We fit the data to the functional form: $A_1 \cos(2\pi f_1 t + \phi_1) e^{-t/\tau_1} + A_2 \cos(2\pi f_2 t + \phi_2)$. As will be justified later, a decay timescale $\tau_1$ is included for the first component only.  The two frequencies extracted from the fit are $f_1$=$2.05\pm0.01$~THz and $f_2$=$2.22\pm0.01$~THz (in energy units: $h f_1 =8.48 \pm 0.04$~meV and $h f_2 = 9.18 \pm 0.04$~meV) with damping term $\tau_1 = 2.7 \pm 0.4$~ps.  The corresponding phases are $\phi_1$ = $  (-0.46 \pm 0.04)\pi$ and $\phi_2$ = $(-0.08 \pm 0.09 ) \pi$.

Before discussing these results we wish to perform a similar analysis on the CB.  Since the CB is a bulk band not associated with sharp spectral features, a similar MDC-fitting analysis cannot be performed.  Instead, in Fig. \ref{oscillations}(c) we plot the transient photoemission intensity  within the integration window shown in the inset.  Because this window is centered on the band edge, its enclosed intensity is sensitive to energetic shifts of the CB.  Again we subtract a smooth background (a 10$^\textrm{th}$ order polynomial) and plot the residual in Fig. \ref{oscillations}(d).  In this case the data is well fit as an undamped oscillation: $A \cos(2\pi f t + \phi)$ with $f$ = $2.23\pm0.01$~THz ($hf = 9.22\pm0.04$~meV) and $\phi$=$(-0.09 \pm 0.03)\pi$.

Due to the similarity of their frequencies, we ascribe the 2.23~THz oscillation observed in the CB and the 2.22~THz oscillation observed in the SS to a common mode origin.  The lack of damping of this mode in the CB is the justification for why no corresponding damping term was included in the fit of the SS oscillations.  The 2.05~THz frequency in the SS is attributed to a mode present only near the surface.  The fitting results  indicate a $\pi/2$ phase shift between the 2.23~THz bulk and 2.05~THz surface oscillations, with the bulk (surface) mode oscillating as a cosine (sine).  Conventionally this cosine-sine distinction is attributed to displacive vs impulsive excitation mechanisms \cite{Dekorsy2000}.

\begin{figure}[bl]
\resizebox{\columnwidth}{!}{\includegraphics{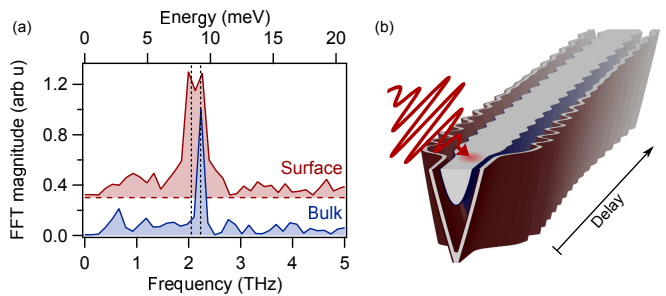}}
\caption{(a) Magnitude of the Fourier transforms of the surface and bulk oscillations. The vertical dashed lines are at 2.05 and 2.23~THz (8.48 and 9.22~meV) .  (b) Cartoon depiction of the optical excitation of coherent phonons in the bulk and surface bands.  For clarity, the magnitude of the oscillations is greatly exaggerated with respect to that observed in the experiment.
\label{FFTandCartoon}}
\end{figure}

In addition to this fitting analysis, we have computed  Fourier transforms of the surface (Fig. \ref{oscillations}(b)) and bulk (Fig. \ref{oscillations}(d)) oscillations, shown in Fig. \ref{FFTandCartoon}(a). It is again clear that there is a shared mode with frequency $\sim$2.23~THz, with an additional frequency 2.05~THz present only at the surface. Fig. \ref{FFTandCartoon}(b) summarizes these findings with a cartoon of the oscillations (the magnitude is exaggerated for clarity). 

We now discuss the physical origin of these oscillations. The A$_\textrm{1g}$ phonon oscillations  observed in TRR have frequencies of 2.09$\sim$2.19~THz, which can be compared with the 2.23~THz oscillation we measured in the CB \cite{Qi2010, Kumar2011, Chen2012,Norimatsu2013a, Glinka2013}.  The fact that our measured frequency is slightly higher may be attributed to lattice stiffening at the measurement temperature of 40~K, as compared to room temperature for the TRR measurements \cite{Kim2012,Irfan2014}.  The appearance of binding energy oscillations  in the CB and SS at this frequency demonstrates that both the bulk and surface electrons couple to this phonon mode \cite{Khan1984}.

The question remains as to why an additional red-shifted frequency is present in the SS.  We attribute this to a softening of the A$_\textrm{1g}$ mode at the crystal surface. The abrupt termination of the crystal at the surface leads to a reduction of the interlayer van der Waals forces which serve as restorative forces for out-of-plane lattice distortions \cite{Zhang2011}.  The corresponding softening of the phonon frequencies is then experienced only by electronic states which are localized near the surface \cite{Eremeev2012}.

\begin{figure}
\resizebox{\columnwidth}{!}{\includegraphics{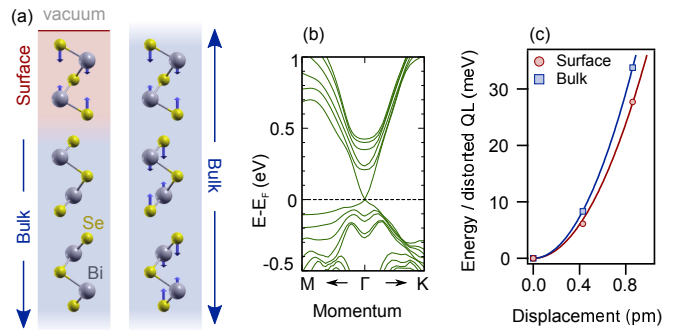}}
\caption{ (a) The lattice distortions employed to compute the surface phonon softening \cite{Kokalj2003}. The arrows represent the A$_\textrm{1g}$ distortions.  The structure on the left represents the upper half of the 6-QL slab used to model surface-only distortions.  A vacuum layer separates each 6-QL slab.  On the right is the 3-QL structure used to model bulk distortions. There is no vacuum layer and hence no surface.    (b) The band structure obtained for the 6~QL slab, which has the Dirac SS clearly visible.   (c)  Total energy of the frozen phonon configurations (per distorted QL) for bulk and surface, showing the softening of the surface phonon.  Solid lines are parabolic fits (without linear terms).
\label{TheoryFigure}}
\end{figure}

To substantiate this understanding we have performed DFT calculations showing the softening of A$_\textrm{1g}$ lattice distortions at the surface as compared to the bulk. The lattice was distorted to mimic the A$_\textrm{1g}$ phonon such that within each quintuple layer (QL), the central Se ion was kept fixed while the Bi and outer Se were distorted.  The Bi displacement was chosen to be 1/2 of the Se displacement.  Two lattice configurations were utilized, shown in Fig. \ref{TheoryFigure}(a). To model bulk distortions we employ a 3-QL system which was relaxed to obtain equilibrium internal coordinates.  To model surface-only distortions we use a 6-QL slab with a 33~\AA $ $ vacuum layer (similarly relaxed to obtain equilibrium coordinates, but keeping the inner 2 QLs fixed at bulk coordinates). Surface A$_\textrm{1g}$ distortions modulate the QLs nearest to both surfaces, but leave the inner 4 QLs fixed. To demonstrate that this structure accurately models Bi$_2$Se$_3$, in Fig. \ref{TheoryFigure}(b) we show the computed band structure of the undistorted 6-QL slab, which exhibits a Dirac surface state within a bulk bandgap.

In Fig. \ref{TheoryFigure}(c) we compute the lattice energy cost $U$ as a function of lattice displacement $\Delta x$ .  We fit the curves assuming a harmonic oscillator-type response ($U \sim \omega^2 \Delta x^2$  where $\omega$ is the oscillation frequency) and  deduce a  $9.7\%$ softening of the A$_\textrm{1g}$ phonon frequency at the surface with respect to the bulk.  This agrees well with the $8\%$ softening observed in the experiment.

Our findings are consistent with a thickness-dependent Raman study on Bi$_2$Se$_3$ nanoplatelets, which similarly found an A$_\textrm{1g}$ frequency red-shift of $\sim6\%$ due to a reduction of restorative van der Waals forces  \cite{Zhang2011}.  We note that similar surface-softening effects have been observed in other materials; for example in Gd(0001) an $\sim18\%$ red-shift was observed \cite{Bovensiepen2004}.

This work showcases the capability of trARPES to resolve distinct modes coupling to individual bands.  Note that while traditional ARPES can only perform self-energy analysis on quasi-2D bands with well-defined dispersions,  here we have resolved EPC on a 3D bulk band.  Furthermore, by measuring  many periods of oscillation we can determine and distinguish mode energies with a resolution of $\sim0.01$~THz $\sim$ $40~\mu$eV,  far exceeding the typical energy resolution of traditional ARPES and even rivaling that of Raman spectroscopy. Finally, we have directly resolved the coupling of the 8.48~meV optical A$_\textrm{1g}$ phonon to the SS band,  which corroborates results from  techniques lacking band sensitivity \cite{Zhu2011,Zhu2012,Howard2013,Costache2014} which suggested that this mode is the dominant scattering channel for Dirac electrons in Bi$_2$Se$_3$.  This provides direct insight into the mechanism limiting electronic conduction in TIs, and therefore has profound implications for future work exploring the potential of these materials for low-dissipation applications.
 

\begin{acknowledgments}
We acknowledge helpful discussions with M. Sentef.  This work was primarily supported by the U.S. Department of Energy, Office of Science, Basic Energy Sciences, Materials Sciences and Engineering Division under contract DE-AC02-76SF00515. J.A.S. and S.-L.Y. acknowledge support from the Stanford Graduate Fellowship. D.L. acknowledges support from the Swiss National Science Foundation. The work by A.F.K. was supported by the Laboratory Directed Research and Development Program of Lawrence Berkeley National Laboratory under the U.S. Department of Energy contract number DE-AC02-05CH11231.  
\end{acknowledgments}

\bibliography{BibTex}

\end{document}